*Costantino Sigismondi*

# Lunar impacts during eclipses separated by a Metonic cycle on Jan 21, 2000 and 2019: a possible origin from daytime Sagittarids/Capriconids meteor shower

Costantino Sigismondi *(ICRA/Sapienza &G. Ferraris InstituteRoma)*
Submitted February 07, 2019,                                    sigismondi@icra.it

**Abstract** The lunar impact claimed by Zuluaga et al. (2019) during the total eclipse of 21 January has been discussed widely by his research group, introducing some results from the technique of gravitational ray-tracing.
A similar event of magnitude 6 was observed visually by the author during the eclipse of 19 years before, that was published under the name of "Padua event" (Sigismondi and Imponente, 2000a,b) and a video was obtained independently by Gary Emerson (Cudnik, 2002) in the US at the same time. The remarkable repetition of such a phenomenon after 19 years deserves some investigation about known active meteor shower on Jan 21 with radiant comprised between the solar longitude 300.7° of January 21 and +/- 60° and declination also departing no more than 60° from the solar one. The amount of 60° is the FWHM of a simple modulated probability model on the visibility of a lunar meteor impact with the cosine of the angle comprised between the line of sight and the normal to the lunar surface. The candidate of this search is the daytime shower Sgr/Cap DSC115 with meteoroid velocities around 26 km/s.

**Sommario** L'impatto lunare osservato da Zuluaga et al. (2019) durante l'eclissi totale del 21 gennaio scorso è stato discusso largamente da quel gruppo di ricerca, con l'introduzione di acluni risultati ottenuti con la tecnica del ray-tracing gravitazionale. Un evento simile di magnitudine 6 è stato osservato dall'autore proprio durante l'eclissi lunare di 19 anni prima, pubblicata con il nome di "evento di Padova" (Sigismondi e Imponente, 2000a,b) e un video era stato ottenuto indipendentemente da Gary Emerson (Cudnik, 2002) negli Stati Uniti allo stesso momento. La notevole ricorrenza di tale fenomeno dopo 19 anni richiede uno studio riguardo le piogge meteoriche note attorno al 21 gennaio, con radianti compresi tra la longitudine solare 300.7° del 21 gennaio e +/-60° e analogamente per quanto riguarda la declinazione rispetto a quella solare. Il valore di +/-60° è ottenuto come larghezza a metà altezza di una semplice distribuzione di probabilità modulata secondo una funzione coseno dell'angolo formato tra la normale alla superficie lunare e la linea di vista. La pioggia meteorica che soddisfa a queste condizioni è la DSC115 Day-time (che avviene solo in pieno giorno) Sagittaridi/Capricornidi, con velocità attorno ai 26 km/s.

**Introduction**

The interest on lunar impacts after gained momentum during the last lunar eclipse of January 20-21 2019, because of an observation videorecorded by Zuluaga et al. (2019) and mirrored worldwide by the social media and the news.
It is remarkable that while the authors tought to be the first to observe an impact during an eclipse, it occurred exactly 19 years before, during another eclipse with the Moon nearly in the same celestial position (Cudnik, 2002). My observation was visual and I located the event in the lower half of the Moon

**Celestial Mechanics: lunar nodes precession, Saros and Metonic cycles**

We know that the lunar nodes precess in 18.6 years, and that after a Saros cycle 18.03 years or 18 years+10 or 11 days and 8 hours another eclipse can occur with the lunar profile oriented in the same way of the previous eclipse. A sequence of eclipses separated by a Saros are grouped into a Saros series.
The eclipses of 21 January 2000 and 2019 belong respectively to the Saros series 134 and 135. (Espenak, 2019)
A Metonic cycle puts togheter lunar phases and solar tropical year, stating that after 19 years the Moon has the same phase in the same day. Less known is the fact that eclipses can repeat also after a Metonic cycle, even if they do not belong to the same "Saros" series. This is due to the fact that either Metonic cycle and Saros are multiple integer of the anomalistic month, respectively 235 and 255.

*Costantino Sigismondi*

## Two bright lunar impacts in the same day of the year, the eclipse condition

In order to identify a possible meteor shower responsible of at least two impacts observed in the same date and separated by exactly 19 years we have to consider that the Moon is nearly 180° from the Sun, and the radiant of the meteor shower should be most probably near the solar position during the eclipse. A probability modulated by the cosine of the angle between the line of sight and the perpendicular to the lunar surface (actually a lunar sphere) has its maximum in the lunar center and decreases to half of this value 60° degrees apart: the radiant of this shower responsible of the impacts should be located +/-60° FWHM either from λ Sun and δ Sun.

## Search of active meteor showers on January 21

The list of all meteor showers available at the Minor Planet Center shows that the DSC 115 Daytime Sagittarids/Capricornids fulfills our requirements.
The mean velocity of the meteors is 26 km/s.
The radiant of 115 DSC is slightly West of the Sun (Rendtel, 2014) i.e. 301°<λ <325° and the velocity of meteoroids is 26 km/s. The parent bodies of the 115 DSC are probably the asteroid 2001 ME1 and the comet 169P/NEAT. This shower is the twin of the 001 CAP, the alpha Capriconids visibile in the Summer in nightime (Sigismondi, 2016 and 2018).

## Conclusions

Zuluaga et al. exclude a meteor shower origin of their impact, also in a private communication (Zuluaga 2019) consider the event of a lunar impact rather common to occur also during an eclipse, and found the velocity of their rogue meteor around 14 km/s, while here it is suggested that this meteor was belonging to a daytime meteor shower, the Sagittarids/Capriconids 115 DSC active in the second part of January, and responsible of the two events in 2000 and 2019 observed both during a total eclipse occurred in the same day with the Moon in the same position of the sky.

In the case of these lunar eclipses two meteoroids of a daytime meteor shower were visible by us exceptionally during the night, because they hit the Moon, at the antipodal position with respect of the Sun, and nearly completely obscured.

**References**

[1] Zuluaga, J. et al. arxiv1901.09573 2019
[2] Espenak, F. http://www.mreclipse.com/ on saros series
[3] Imponente, G. and C. Sigismondi, 2000astro.ph..6210I
[4] Imponente, G. and C. Sigismondi, WGN 28, 54 2000
[5] Imponente, G. and C. Sigismondi, WGN 28, 230 2000 .
[6] Rendtel, J. Proc. of the IMC Meeting, Giron, 2014, p. 93-97
[7] Cudnik, B. The Strolling Astronomer, 44, 7 2002
[8] Minor Planet Center list of all meteor shower www.ta3.sk/IAUC22DB/MDC2007
[9] Sigismondi, C., The dynamical history of Piscis Austrinids meteor shower: rare mid-summer comet's remnants, Proc. of Rosetta Meeting, Padova (2018)
[10] Sigismondi, C., Gerbertus 10, 57 2016